\documentclass[12pt,preprint]{aastex}
\input epsf.tex
\begin{document}
 \title{A Spitzer Search for Infrared Excesses around Massive Young White Dwarfs\altaffilmark{1}}
 \author{Brad M. S. Hansen\altaffilmark{2,3}, Shri Kulkarni\altaffilmark{4} \& Sloane Wiktorowicz\altaffilmark{5}}
\altaffiltext{1} {Based on observations made with the Spitzer Space Telescope}
\altaffiltext{2}{Department of Astronomy, University of California Los Angeles, Los Angeles, CA 90095, hansen@astro.ucla.edu }
\altaffiltext{3}{Alfred P. Sloan Research Fellow}
\altaffiltext{4}{McArthur Professor of Astronomy and Planetary Sciences, California Institute of Technology, Pasadena, CA, 91125, srk@astro.caltech.edu}
\altaffiltext{5}{Department of Geological and Planetary Sciences, California Institute of Technology, Pasadena, CA, 91125, sloane@gps.caltech.edu}

 \slugcomment{\it 
}

\lefthead{Hansen, Kulkarni \& Wicktorowicz}
\righthead{MWD IRXS}

\begin{abstract}
We examine 14 hot white dwarfs for signs of infrared excess using the Spitzer Space
Telescope. Twelve of the objects are massive white dwarfs which have been suggested
to be the result of binary mergers. The remaining two objects are undermassive white
dwarfs which again may be the result of mergers or the inspiral of a substellar
companion. In no case do we find any evidence for significant infrared excesses out to 
wavelengths of $8 \mu m$. This places strong constraints on the presence of orbiting
dust and weaker constraints on the presence of close substellar companions.
\end{abstract}

\keywords{white dwarfs -- planetary systems: protoplanetary disks -- infrared: stars}

\section{Introduction}

The mass distribution of white dwarf stars is one of the fundamental
measures of stellar evolution. The majority of white dwarfs have
masses distributed in a narrow range between $0.5-0.6 M_{\odot}$ (Weidemann \& Koester 1984;
Bergeron, Saffer \& Liebert 1992 --BSL; Liebert, Bergeron \& Holberg 2005 --LBH). As studies
of this type grew steadily more accurate, two additional, smaller peaks appeared
in the distribution (BSL, LBH). The first is centered around a mass $\sim 0.4 M_{\odot}$ and
represents the outcome for stars in close binaries, whose stellar evolution has been
truncated by Roche lobe overflow before they reached the Helium flash. The second minor
peak lies at higher masses $> 0.8 M_{\odot}$. This excess of massive white dwarfs is even
more extreme amongst white dwarfs selected from ultraviolet or X-ray surveys (Marsh et al 1997;
Vennes et al 1997) because surveys biased towards hotter stars are more likely to find
massive white dwarfs (a result of their smaller radii).

This apparent excess of massive white dwarfs is of interest because it
represents a potential sample of white dwarf merger remnants. A natural outcome of
stellar binary evolution is a significant population of double degenerate
white dwarf binaries in close orbits, close enough that they should merge within a
Hubble time (Webbink 1984; Iben 1990; Yungelson et al 1994). Most of these binaries
have total masses well below the Chandrasekhar mass limit and so are expected to
form a single isolated remnant star. The expected population of $>0.8 M_{\odot}$
degenerate merger remnants is thus quite naturally associated with the observed
excess of massive white dwarfs. However, there is as yet no convincing evidence 
that any of these massive white dwarfs are indeed the result of a merger.

Numerical simulations of the merger of two unequal mass white dwarfs suggest 
that the less massive of the pair is tidally disrupted and accreted onto the
more massive one in the form of a thick accretion disk (Mochkovitch \& Livio 1989;
Benz et al 1990; Guerrero, Garcia-Berro \& Isern 2004). Thus, it would appear
that the natural outcome of such a picture is a hot white dwarf surrounded by
a thin remnant disk which contains most of the angular momentum. The nature of this
disk is likely to be very unusual, given that white dwarfs are composed primarily of
Carbon and Oxygen, rather than the Hydrogen and Helium of more traditional circumstellar
disks. The
observational signature of such circumstellar disks is an infra-red excess, as dust particles
in the disk absorb stellar photons and re-radiate them at longer wavelengths characteristic
of the cooler temperatures in the disk. Indeed, just such an excess is observed for 
the white dwarf G29-38 (Zuckerman \& Becklin 1987), although the source of the reprocessing
disk is likely to be something else in that case (Graham et al 1990; Debes \& Sigurdsson 2002; Jura 2003).
 Thus, as a test of the merger hypothesis for the
origin of massive white dwarfs, we have undertaken a Spitzer program to search for infra-red
excesses around the youngest/hottest massive white dwarfs.
 
\section{Observations}

The sample was selected from those white dwarfs found in the EUVE and ROSAT surveys,
which were optically classified as being massive ($>1 M_{\odot}$). Selection from
these surveys guarantees that the white dwarfs are hot and therefore young (effective
temperatures $>35,000$K correspond to ages $<10^7$ years). This strategy has its strengths and
weaknesses. It is observed that infra-red excesses around main sequence stars disappear
on timescales of a few $\times 10^8$~years (Habing et al 1999; Rieke et al 2005, and
references therein), so that we placed a great emphasis on observing stars in
the earliest possible stages of their present incarnation. On the other hand, LBH
suggest that, while there is definitely an excess in massive white dwarfs over that
expected from single star evolution, 
the UV/X-ray samples may be biased towards the progeny of nearby O/B associations
which may dilute the fraction of merger remnants. We shall return to this subtle point
in \S~\ref{Discuss}.

Drawing from the EUVE (Vennes et al 1997) and ROSAT (Marsh et al 1997) samples, we
chose twelve white dwarfs with masses $>1 M_{\odot}$ (based on the optical spectral
classifications). To this sample of potential merger remnants we also added two
low-mass white dwarfs which have also been suggested as merger remnants (Maxted \& Marsh 1998)
because they show no evidence for a companion. Table~\ref{BigTab1} shows the resulting
sample of fourteen objects, with temperatures, masses and distances.

In order to determine the appropriate bandpass, let us consider briefly the
kind of remnant disk one might expect from a merger remnant. White dwarfs are
more compact than main sequence stars and the angular momentum deposited in
the disk is only sufficient to produce disks of extent $\sim 10^{11}$cm.
As such, we
expect the disks to be somewhat hotter than traditional protoplanetary disks. Simple models suggest that the
emission from a passively reprocessing disk will peak at wavelengths $\sim 10 \mu m$. 
Thus, observations were planned in the Spitzer IRAC bandpasses. The
spectral coverage of Spitzer from 3.6--8$\mu m$ is well-suited to the expected spectral
shape. These issues will be discussed further in \S~\ref{models}.

The Spitzer observations were taken in all four IRAC bandpasses, with 3 frames of
 12s exposures in a 5 dither pattern. The resultant total exposures were 180s for
each band. The exposure lengths were chosen to produce detections of simple reprocessing
models at levels $\sim 100 \sigma$. The exposures were also sufficiently long to
ensure confident detection of the white dwarf photospheric emission at 3.6$\mu m$, 
allowing us to place stringent constraints in the event of no detection.

\section{Results}

Table~\ref{BigTab2} shows the observed fluxes. We include V-band magnitudes from
the literature and have either measured 2MASS J-band fluxes or culled
results from the literature. There is no convincing
evidence of an infra-red excess in any of the sources. In several cases, the 
observations are good enough to trace the photospheric emission all the way to
8$\mu m$ and in all cases the photosphere is detected at expected levels in the
shortest IRAC bandpass.

\begin{figure}
\begin{center}
\leavevmode
{Due to archive space restrictions, all images are in jpeg
rather than postscript format. The paper may be restored to
original form by converting the jpegs back to postscript and
uncommenting the `includegraphics' line in each of the relevant
figures in the source code file.}
{\bf Figure 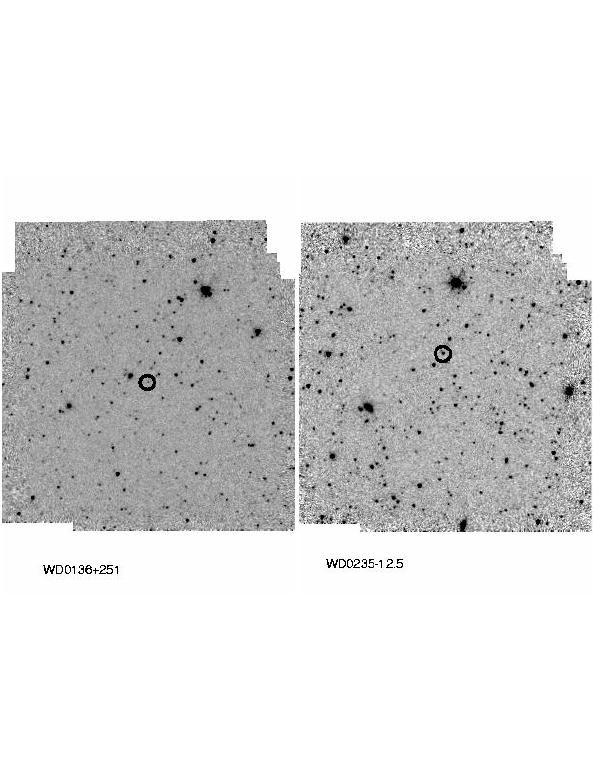 goes here}
\end{center}
\caption{ The circle indicates the location of the white dwarf in
each of these fields.
Both images were taken in the IRAC~2 band.
Each image is approximately $6' \times 6.4'$.
}
\label{chart1}
\end{figure}

\begin{figure}
\begin{center}
\leavevmode
{\bf Figure 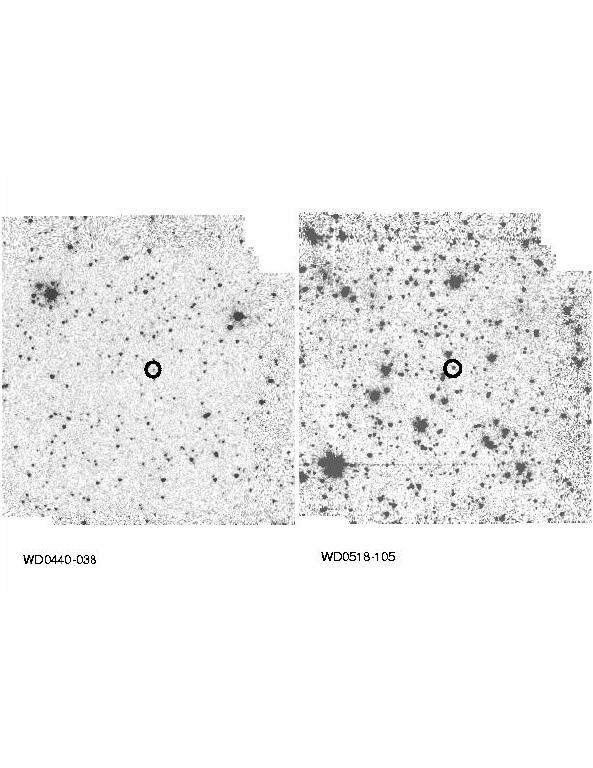 goes here}
\end{center}
\caption{ The circle indicates the location of the white dwarf in
each of these fields.}
\label{chart2}
\end{figure}

\begin{figure}
\begin{center}
\leavevmode
{\bf Figure 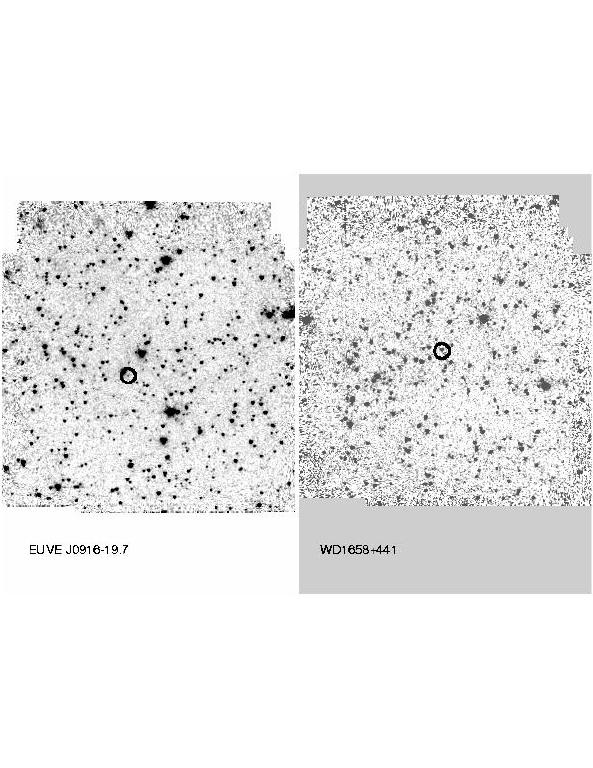 goes here}
\end{center}
\caption{ The circle indicates the location of the white dwarf in
each of these fields.}
\label{chart3}
\end{figure}

\begin{figure}
\begin{center}
\leavevmode
{\bf Figure 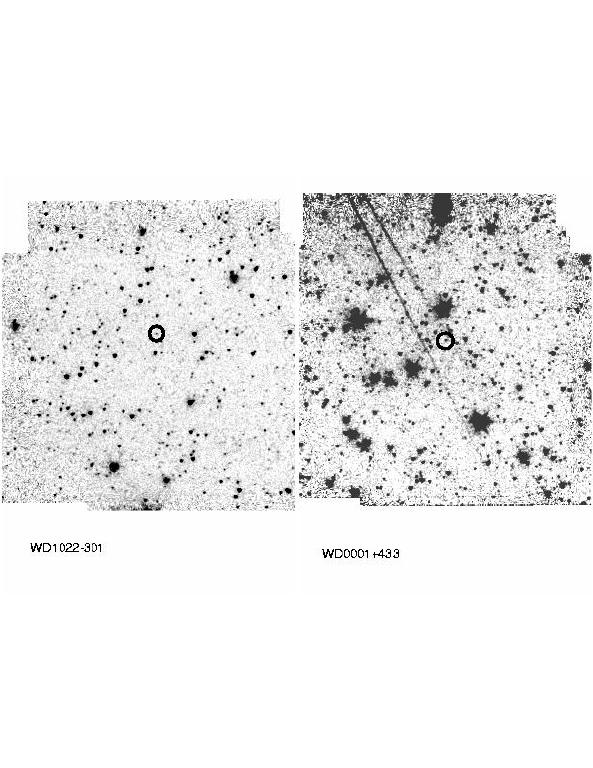 goes here}
\end{center}
\caption{ The circle indicates the location of the white dwarf in
each of the fields.
}
\label{chart4}
\end{figure}

\begin{figure}
\begin{center}
\leavevmode
{\bf figure 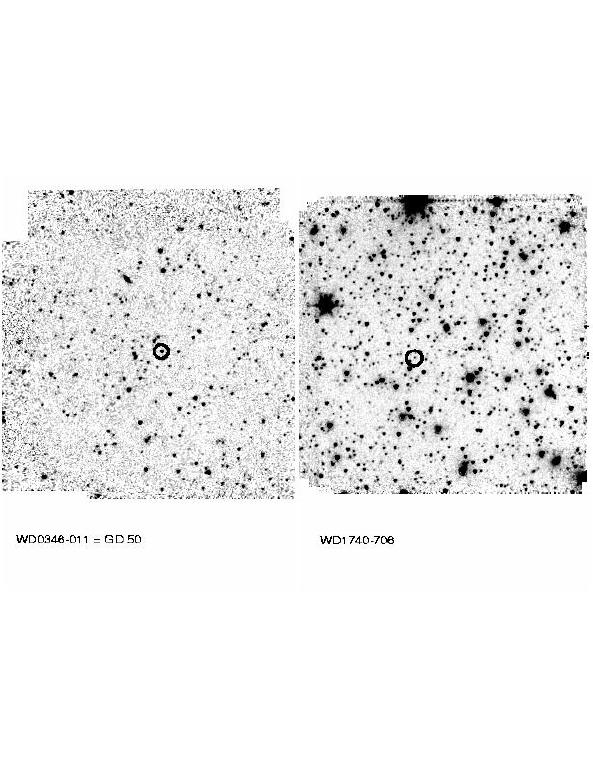 goes here}
\end{center}
\caption{ The circle indicates the location of the white dwarf in
each of the fields.
}
\label{chart5}
\end{figure}

\begin{figure}
\begin{center}
\leavevmode
{\bf figure 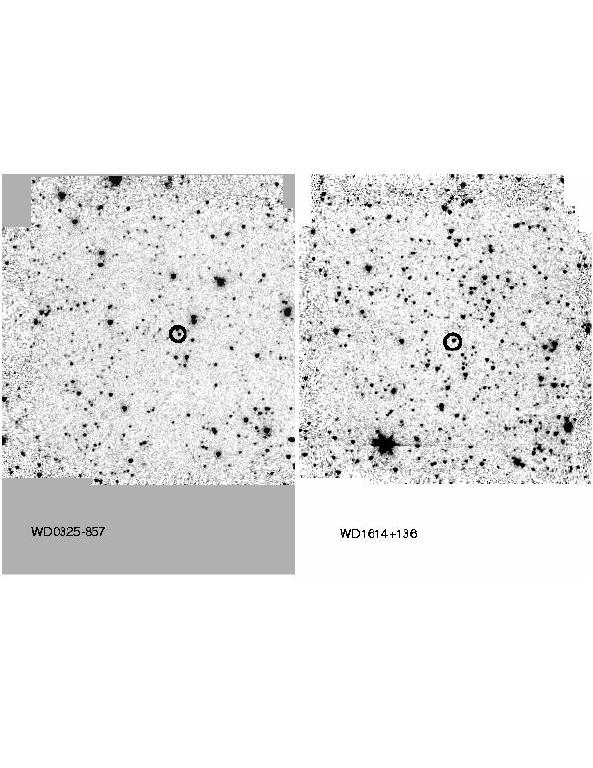 goes here}
\end{center}
\caption{ The circle indicates the location of the white dwarf in
each of the fields.
}
\label{chart6}
\end{figure}

\begin{figure}
\begin{center}
\leavevmode
{\bf figure 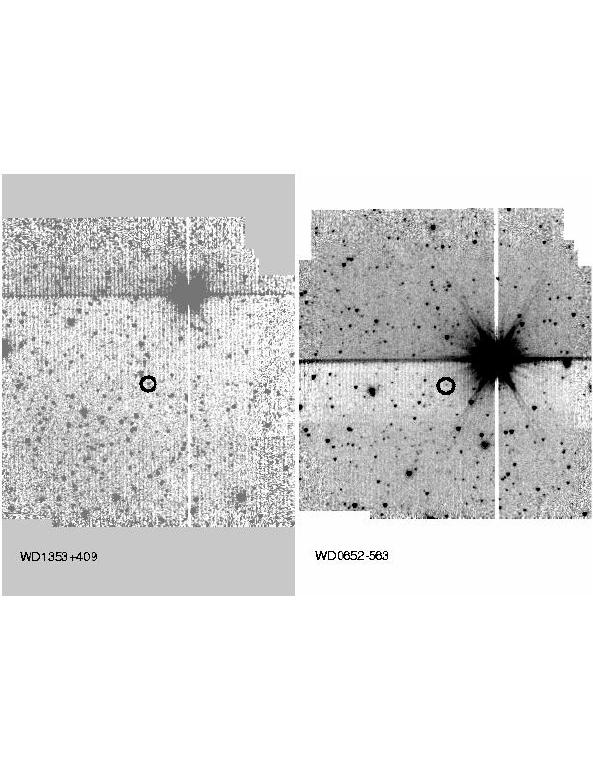 goes here}
\end{center}
\caption{ The circle indicates the location of the white dwarf in
each of the fields.
}
\label{chart7}
\end{figure}

 Matching finding charts between optical and IRAC bands can prove challenging,
especially for objects as blue as ours. Thus, we provide IRAC finding charts for our
objects in Figures~\ref{chart1}--\ref{chart7}.

\subsection{Noteworthy Objects}

\subsubsection{GD~50}

The prototypical supermassive white dwarf, GD~50 was classified as such
by Bergeron et al (1991) on the basis of the anomalous width of the
Balmer lines in the spectrum. Bergeron et al derived $T_{eff} = 43500\pm 1500$~K
and $\log g = 9.00 \pm 0.15$ (both 3$\sigma$ errors), yielding a mass $\sim 1.2 M_{\odot}$. 
They noted that, under the assumption that massive white dwarfs are born from massive
stars, one would expect the total stellar age of this star to be short ($\sim 10^7$~years)
so that it was rather surprising to find no evidence of recent star formation in
the vicinity. Following this line of argument, one may consider GD~50 one of the
most hopeful candidates for finding evidence of a recent merger. 

\begin{figure}
\begin{center}
\leavevmode
\includegraphics[width=14cm,angle=0]{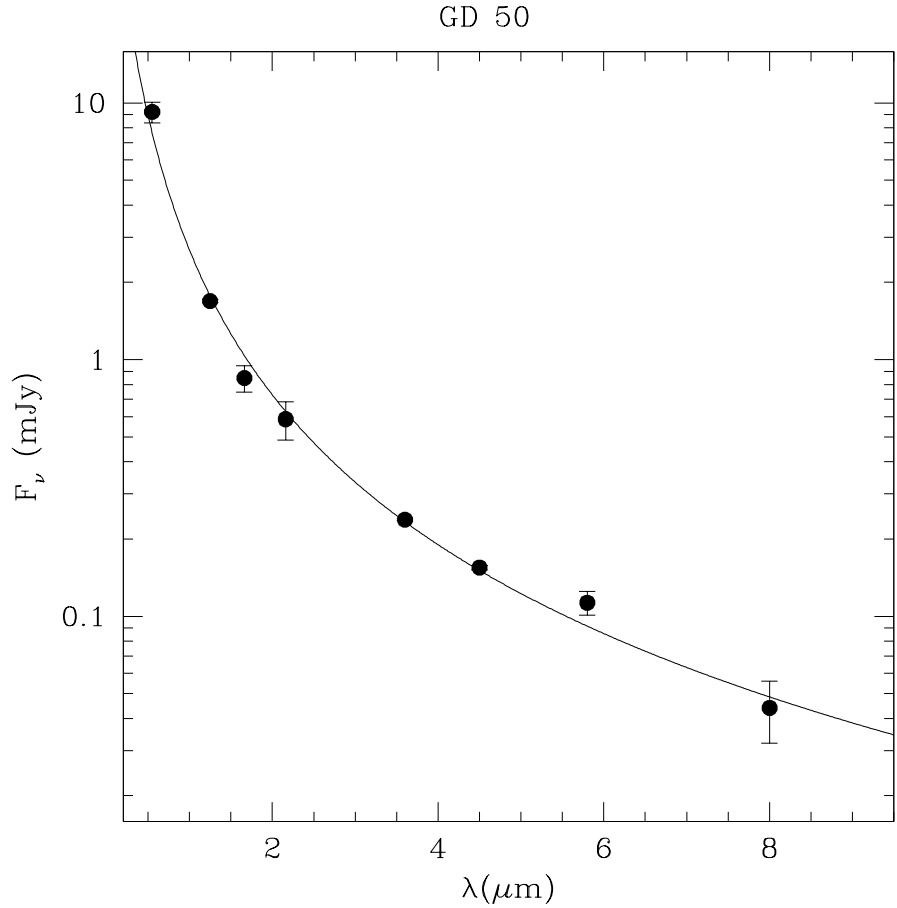}
\end{center}
\caption{ The points show flux measurements in the
four IRAC bands, the three 2MASS bands and in the V band.
The solid line is for a 40,000~K black body.
There is no evidence of an infrared excess.
}
\label{GD50_flux}
\end{figure}

The white dwarf is solidly detected in the
first three IRAC bandpasses and  marginally detected in the fourth bandpass.
The resulting photometry is shown in Figure~\ref{GD50_flux}. We see that
although the object is detected in the optical, in 2MASS and by Spitzer,
all flux measurements are consistent with photospheric emission. We show
 a 40,000~K black body as a solid line.
Matching the model fluxes to the observed fluxes, assuming the
spectroscopic temperature,  provides a constraint on
the ratio of radius ($R_9$ in units of $10^9$cm) to distance ($D_{pc}$ in
units of parsecs)
$$
\frac{R_9}{D_{pc}} = 0.0108 \pm 0.0016
$$
which is in excellent agreement with the values quoted in Bergeron et al
($R=0.0057 R_{\odot}$ and $D\sim 37$pc, which imply $R_9/D_{pc} = 0.011$).
Thus, our measurements of GD~50 are consistent with its status as a massive
white dwarf if the distance is of order the nominal 37~parsecs.

\subsubsection{EUVE J0916-19.7}

\begin{figure}
\begin{center}
\leavevmode
{\bf Figure 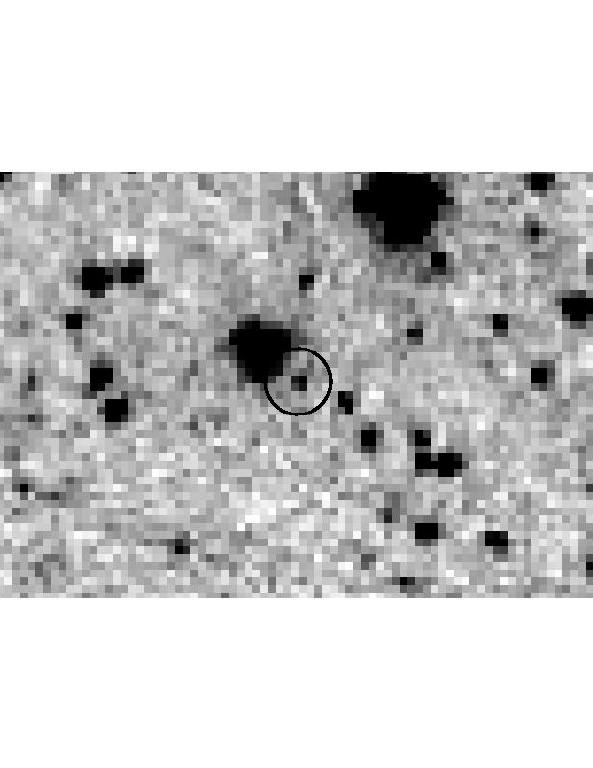 goes here}
\end{center}
\caption{ The circle indicates the location of the white dwarf EUVE0916-19.7 in
the IRAC~1 bandpass. The star to the upper left is most
likely the source of the infra-red excess reported by Green et al., as shown
in Figure~\ref{Flux7}}
\label{0916}
\end{figure}

\begin{figure}
\begin{center}
\leavevmode
{\bf Figure 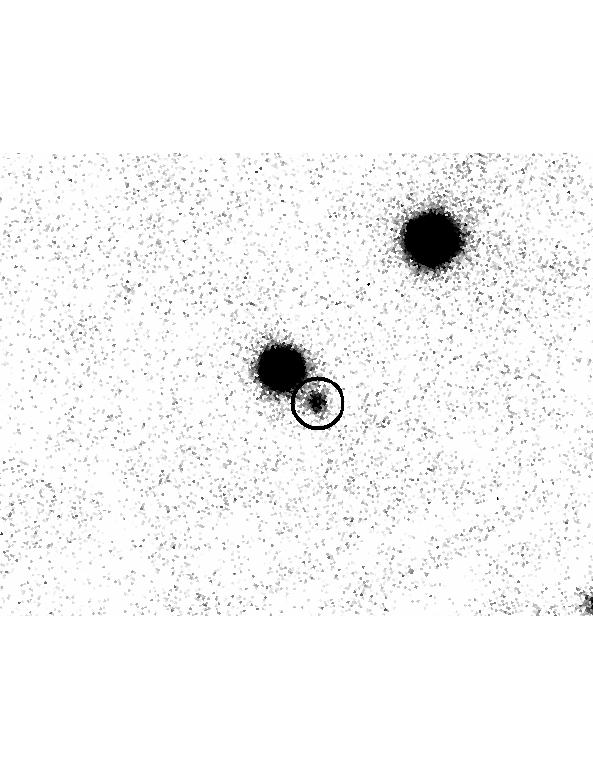 goes here}
\end{center}
\caption{ The circle indicates the location of the white dwarf EUVE0916-19.7.
This is an I-band image taken with the Palomar 60$''$ telescope.
}
\label{0916_2}
\end{figure}

\begin{figure}
\begin{center}
\leavevmode
\includegraphics[width=14cm,angle=0]{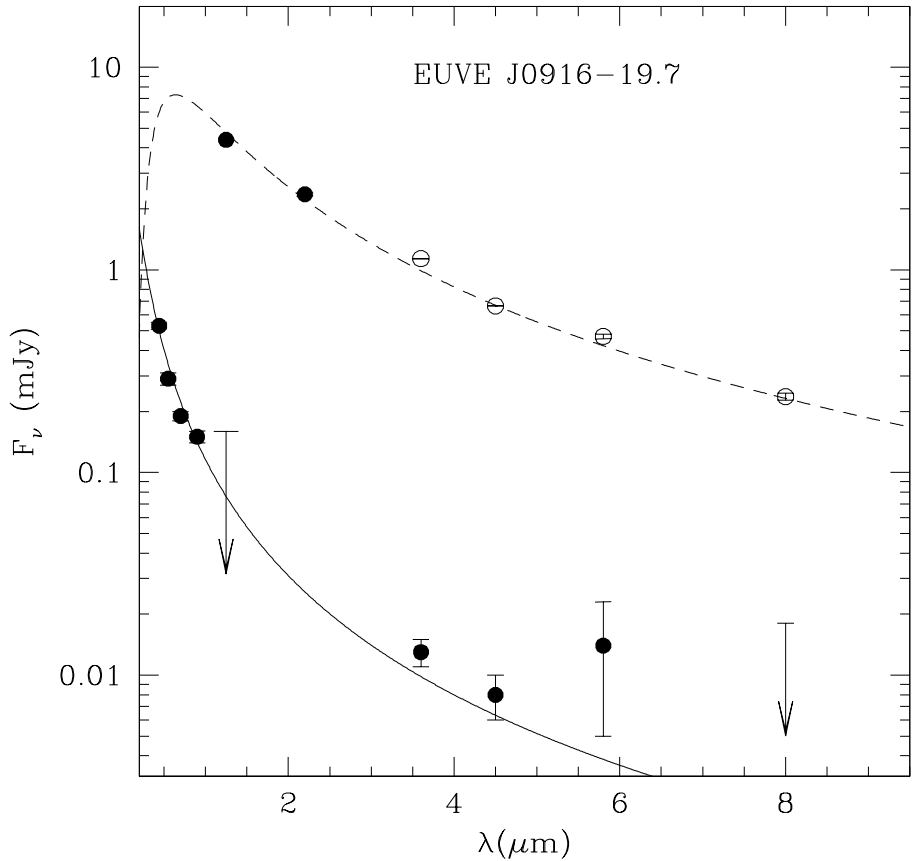}
\end{center}
\caption{ The lower set of points are the flux measurements (4 IRAC
bands and the BVRI magnitudes)
for EUVE J0917-19.7 (the circled source in Figure~\ref{0916}). 
The two solid
circles near the top left are the J and K fluxes Green et al claim
to measure for EUVE J0917-19.7. The four open circles are the IRAC
fluxes of the object 7.2$''$ away from the center of the circle in Figure~\ref{0916}. 
The dashed curve is a black body of temperature 8000~K, while the solid
curve is for a blackbody of 56,400~K.
}
\label{Flux7}
\end{figure}

Green et al (2000) claim a significant near-infrared excess for this object,
with J and K values approximately 4 magnitudes above the values predicted
from atmosphere models. 
As can be seen from Figure~\ref{0916}, they have almost certainly reported the
magnitude for the star $\sim 7.2''$ away from the true source. The circled white
dwarf has flux measurements consistent with photospheric values. To confirm this,
Figure~\ref{0916_2} shows an I-band image of the field taken with the Palomar~60$''$ telescope.
The white dwarf is circled.
 Figure~\ref{Flux7}
shows the near-IR fluxes for both the white dwarf and the nearby star.

We can determine $R_9/D_{pc}$ for these objects as well. The red object
is probably a main sequence star at some distance behind the white dwarf.

\subsubsection{REJ0317-853}

This object has been suggested as a merger remnant on the basis of
several lines of argument. Not only is it massive, but it is also
magnetised and has a rapid rotation period of 725s. All these features
can be plausibly explained by a double degenerate merger (Barstow et al 1995;
Ferrario et al 1997).

\begin{figure}
\begin{center}
\leavevmode
{\bf Figure 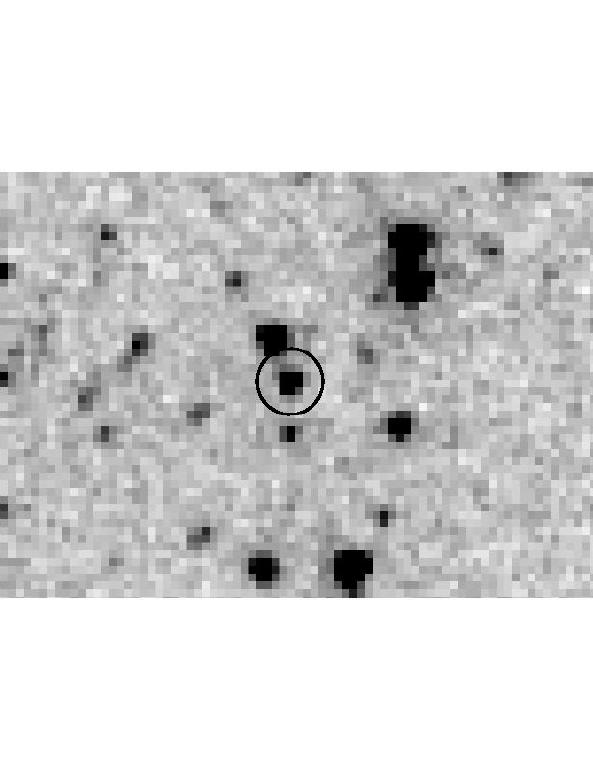 goes here}
\end{center}
\caption{ The circle shows the location of the white dwarf REJ0317-853
in the IRAC~2 passband. The star to the upper left is also a white 
dwarf, LB~9802.
}
\label{Tile}
\end{figure}

\begin{figure}
\begin{center}
\leavevmode
\includegraphics[width=14cm,angle=0]{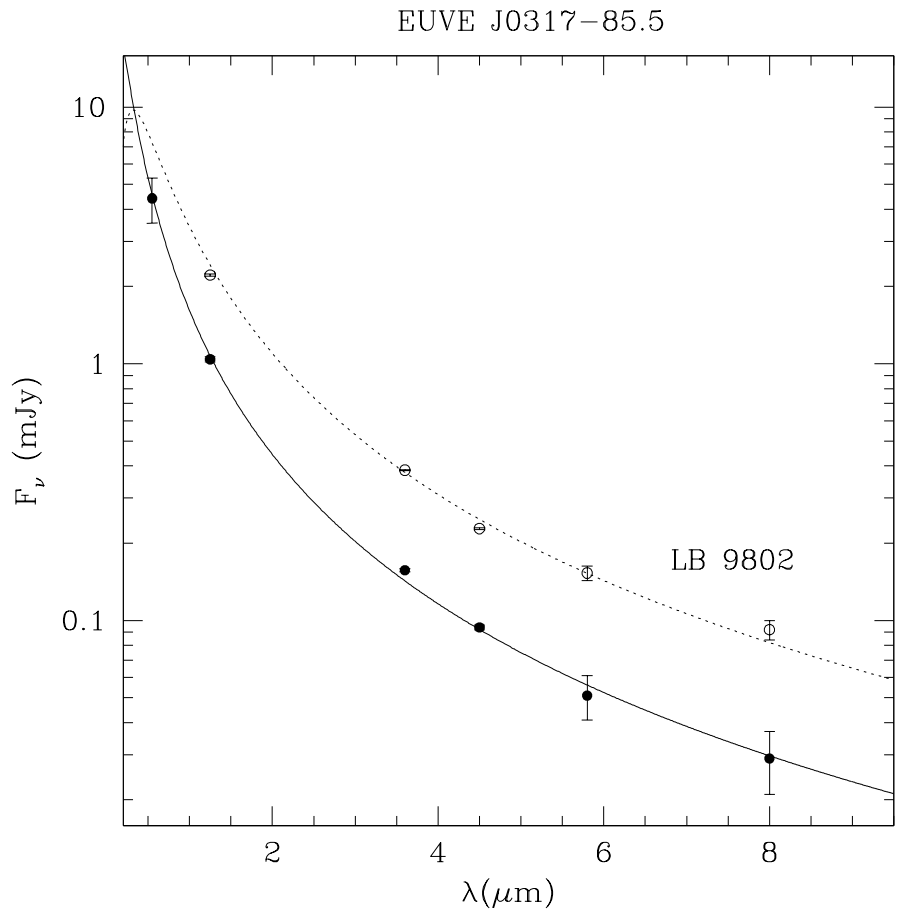}
\end{center}
\caption{ The solid points and curve are the flux measurements for EUVE J0317-85.3
while the open points and dotted curve are for the nearby white dwarf LB~9802.
}
\label{Flux5}
\end{figure}

Figure~\ref{Tile} shows the resulting field. Once again the object is
easily detected in the first three bands and marginally detected in
the longest bandpass. However, all the detections are consistent with
photospheric emission from a 40,000~K white dwarf. The results are shown
in Figure~\ref{Flux5} along with the flux measurements for the close
white dwarf companion LB9802. This white dwarf has a spectroscopic
temperature and gravity ($T_{eff} = 16030\pm 230$, $\log g = 8.19 \pm 0.05$ -- Barstow
et al 1995). It was reported as being at a separation of 16$''$ by Barstow et al,
but the Spitzer image yields a separation of $7.4''$ (a similar number is
obtained from 2MASS).

The temperature determination for EUVE J0317-85.3 is $T_{eff} \sim 40,000$~K, but
there is no spectroscopic gravity. However, Barstow et al argued for the binarity
of the two white dwarfs (on the basis that their close separation was unlikely to
be random). Placing the two objects at the same distance then implies a high mass
for the hotter star because it is fainter. Our Spitzer photometry yields a
similar answer. In the Rayleigh-Jeans tail, the flux ratio (assuming the same distance)
is 
$$
 \frac{F_1}{F_2} = \left(\frac{R_1}{R_2} \right)^2 \frac{T_1}{T_2} = 2.5
$$
if we designate LB~9802 as star 1 and EUVE J0317-85.3 as star~2. Using the
measured temperatures, this implies $R_1/R_2 \sim 2.5$. So, for $R_1 \sim 8 \times 10^8$cm 
(based on the measured gravity of LB9802),
this yields $R_2 \sim 3.2 \times 10^8$cm.

\subsubsection{WD0235-12.5}

\begin{figure}
\begin{center}
\leavevmode
{\bf Figure 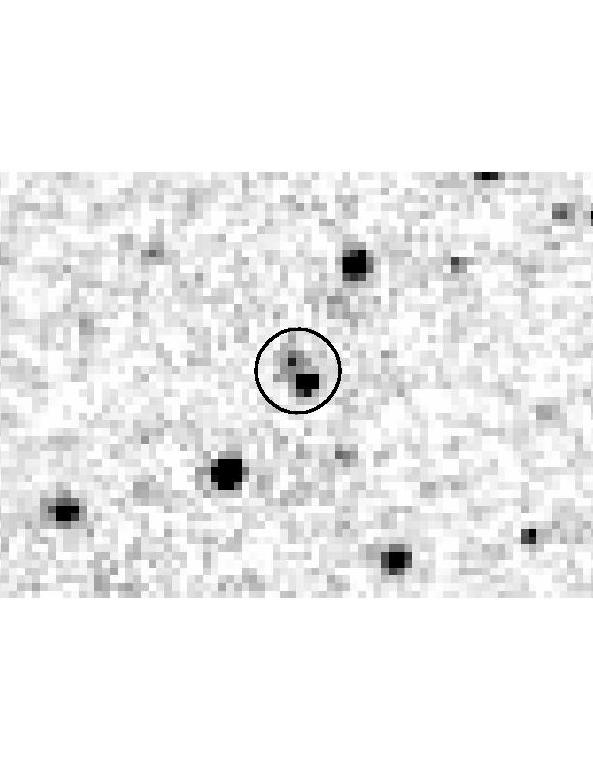 goes here}
\end{center}
\caption{ The circle shows the position of WD0235-12.5 and the neighbouring
object, as viewed in the IRAC~2 (4.5$\mu$m) passband. The separation is
approximately 3 pixels ($\sim 3.6''$).
}
\label{FindComp}
\end{figure}

\begin{figure}
\begin{center}
\leavevmode
\includegraphics[width=14cm,angle=0]{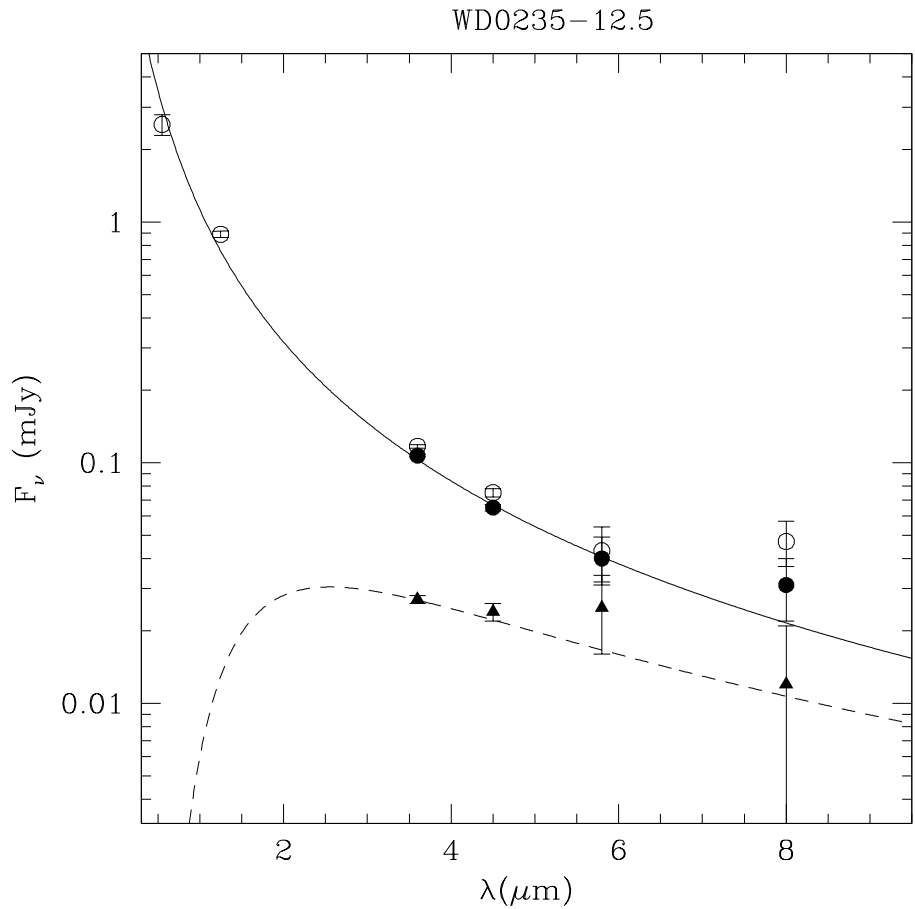}
\end{center}
\caption{ The open points are the flux measurements for WD0235-125 taken
with a standard 3 pixel aperture (to maintain consistency with the bulk
of our sample). The solid circles indicate the flux measured with a
2 pixel aperture and the filled triangles show the flux measured for
the companion. The solid line shows a blackbody of 32,400K.
The dashed line shows a blackbody of 5000~K.
}
\label{Flux2}
\end{figure}

Figure~\ref{FindComp} shows the field around WD~0235-125. Although once again
no credible excess is detected (see Figure~\ref{Flux2}), there is an additional
source within 4$''$ of the white dwarf. Any claim about close companions in fields
as crowded as these must be viewed with caution, but the characteristics of this
object are of interest. The fluxes shown in Figure~\ref{Flux2} are
consistent with a blackbody of $\sim 5000$K and radius approximately twice as
large as that of WD~0235-125 (if the two are physically associated). The separation
of $3.6''$ at a systemic distance of 6.6~pc means an orbital separation of 238~Au. 
This would make it the second star in our sample to have a white dwarf 
companion. However, the cooling age of such a companion is $\sim 3 \times 10^9$ years
(as opposed to $\sim 10^8$ years for WD~0235-125). Perhaps the massive white dwarf is the
product of mass transfer and binary evolution. One problem with this speculation is that
a main sequence star with lifetime
$\sim 3$~Gyr has a mass $\sim 1.4 M_{\odot}$, so that the final white dwarf 
is of comparable mass to the putative progenitor.

\subsubsection{WD1614+136 \& WD1353+409}

These two objects are different from the others in our sample. Identified as low
mass Helium core white dwarfs, these stars are expected to have a close, dark companion
which is responsible for the stripping of the progenitor envelope before the
core mass got large enough to undergo Helium burning. Yet
Marsh, Dhillon \& Duck (1995)
found no evidence for the radial velocity variations that characterise most of
the stars in this class. Possible alternatives to the traditional route involve
either the merger of two low mass white dwarfs (although Marsh \& Maxted 1998 find normal rotation velocities)
 or perhaps stripping by a planetary
mass companion (Nelemans \& Tauris 1998).

\begin{figure}
\begin{center}
\leavevmode
\includegraphics[width=14cm,angle=0]{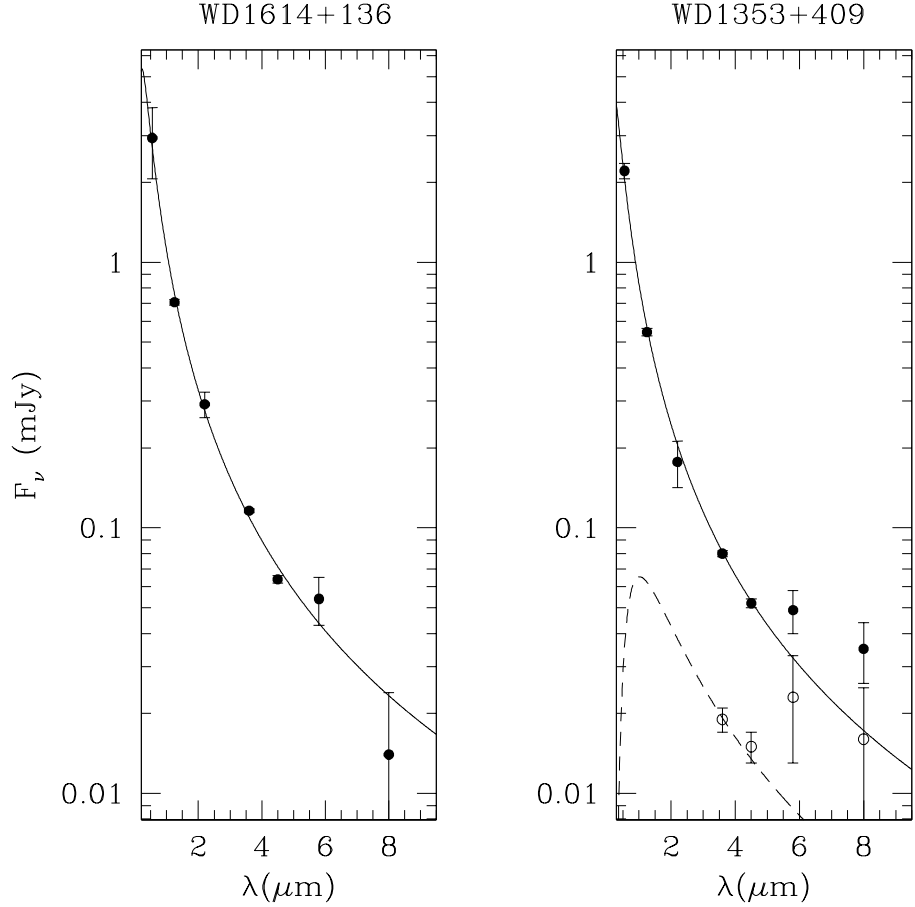}
\end{center}
\caption{The left panel shows WD1614+136 (compared to a 22400~K black body)
and the right-hand panel shows WD1353+409 (compared to a 23600~K black body)/
The dashed curve in the right-hand panel shows a 5000~K black body fitted to
the point source $\sim 5''$ from the white dwarf. Once again, solid points
indicate photometry measured in a 3 pixel aperture and open circles in
a two pixel aperture.
}
\label{Flux6}
\end{figure}

\begin{figure}
\begin{center}
\leavevmode
{\bf Figure 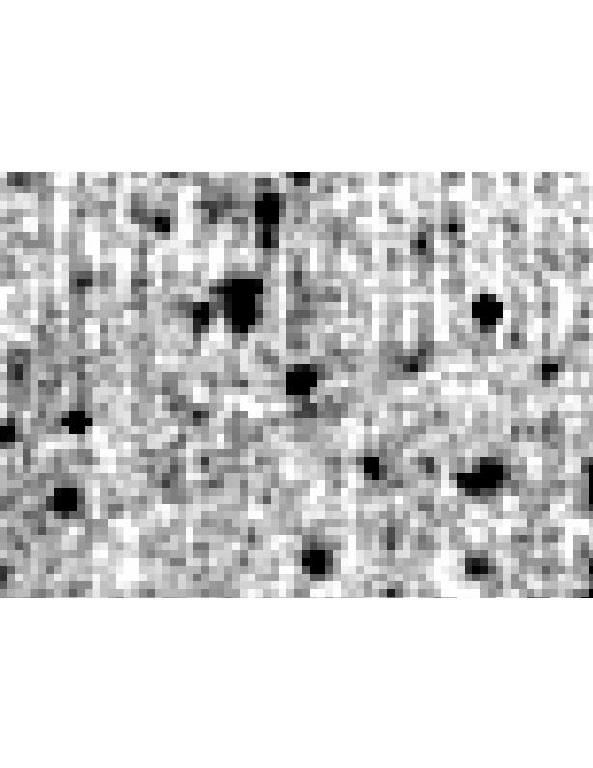 goes here}
\end{center}
\caption{This shows the zoom in around WD1353+409. The presence of a
bright nearby star has caused some banding in this (IRAC1) image, but
the star is still well-measured. Furthermore, the image reveals
a marginal detection of another
 object $\sim 4$ ( $\sim 5''$) pixels away.
}
\label{1353}
\end{figure}

Figure~\ref{Flux6} shows the fluxes measured for WD1614+136 and WD1353+409. Once again
the emission is completely consistent with the photospheric expectations,
from optical wavelengths to 8$\mu m$. The field of WD1353+409 (Figure~\ref{1353}) shows
a marginal detection of a nearby source that may be a companion. If it
is indeed
physically associated, then the flux ratio is consistent with a 5000~K white dwarf
companion. Larger radius objects such as brown dwarfs or planets require lower temperatures
which are inconsistent with the photometry (see \S~\ref{PlanetComps}).

\section{Models}
\label{models}

Given that all of our observations are consistent with photospheric emission,
what upper limits can we place on the presence of a disk?

Despite the fact that conditions in merger remnant disks are far from 
the usual situation, some models do exist (Livio, Pringle \& Saffer 1992; Menou, Perna \& Hernquist 2001; Livio, Pringle \& Wood 2005).
For metallic materials, once the disk temperatures decrease below $\sim 10^4$K
the material will recombine and one might expect the formation of dust. 
Simple models of viscous disk evolution in a merger scenario (Menou et al 2001) suggest that 
the outer edge of the disk can evolve to scales $\sim 4 \times 10^{11}$cm 
before the entire disk becomes neutral. Furthermore, setting the temperature
of the passive disk equal to $10^4$K yields the inner edge of any plausible
disk (interior to this the stellar irradiation can keep the material ionized
even in the absence of viscous heating). This yields a value
$$
a_i \sim 4 \times 10^9 cm \left( \frac{T_*}{45000 K} \right)^{4/3}.
$$
Thus a plausible range of disk scales ranges from $\sim 4 \times 10^9$cm to $4 \times 10^{11}$cm.

\subsection{Flat Black Body Disk}

The simplest disk is a flat, optically thick, passively reprocessing disk in which
each annulus radiates as a black body (Adams et al 1987 and references therein).
We note that this is a more conservative model than that of Livio et al (2005),
which flares as a result of vertical hydrostatic equilibrium. So our constraints
on that model would be even stronger than the ones derived below.

In the flat disk model, the disk luminosity is
$$
L_{\nu} = 8 \pi^2 \cos i \int_{a_i}^{a_o} a B_{\nu}(T(a)) da
$$
where $a_i$ and $a_o$ are the inner and outer radii of the disk and $i$ is
the inclination angle of the line of sight. The
temperature profile $T(a)$ is determined by equating the incident and
radiated luminosity at each annulus,
$$
T = \left( \frac{2}{3 \pi} \right)^{1/4} \left( \frac{R_*}{a} \right)^{3/4} T_*
$$
where $R_*$ and $T_*$ are the stellar radius and effective temperature
respectively. This results in an expected flux
$$
F_{\nu} = 17 \, {\rm mJy} \left( \frac{\lambda}{3.6 \mu m} \right)^{-1/3}
\left( \frac{T_*}{45000 K} \right)^{8/3} \left( \frac{D}{100 pc} \right)^{-2}
\cos i \int_{z_i}^{z_o} \frac{z^{5/3} dz}{e^{z}-1}
$$
for a disk around a star at 100~pc distance ($z = hc/\lambda k T$).
 We have chosen numbers appropriate
to hot white dwarfs rather than the usual main sequence star parameters. We
note that the only tunable parameters for this simple model are the inner and
outer radius and the inclination.

\begin{figure}
\begin{center}
\leavevmode
\includegraphics[width=14cm,angle=0]{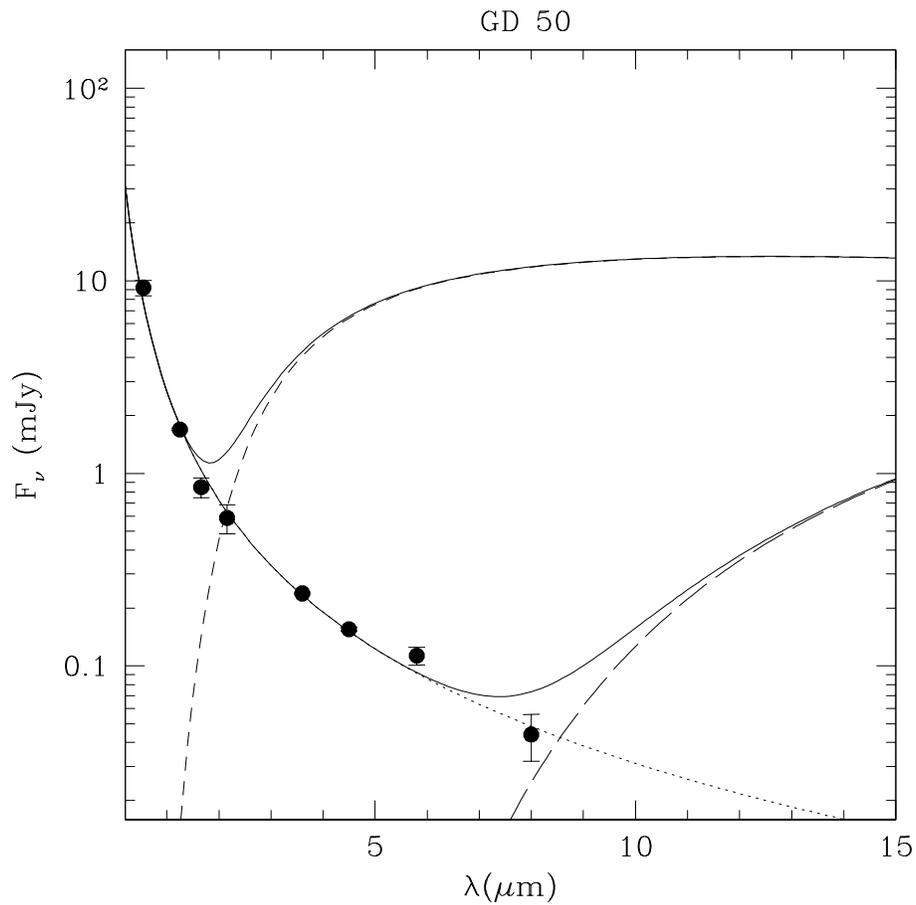}
\end{center}
\caption{ The solid points show the flux measurements for GD~50. The dotted
curve is a black body of the appropriate temperature. The dashed curves
show the emission from a flat, passively reprocessing black body disk around
GD~50. The short dashed curve assumes an inner disk radius of $a_i=10^{10}$cm
and the long dashed curve has $a_i=10^{11}$cm. In both cases we assume an
inclination of 85$^{\circ}$ to the line of sight and outer radius $a_o=10 a_i$.
}
\label{Comp1}
\end{figure}

As can be seen from Table~\ref{BigTab2}, our observations are all at the level
of a fraction of a mJy, so well below the predicted level. Essentially all of
the reasonable parameter space is ruled out. Figure~\ref{Comp1} shows an 
example of the model comparisons with the data for GD~50. In order to fit
the data at the faint end, the inner edge of the disk has to be placed at
$\sim 10^{11}$cm (close to the {\bf outer} edge expected in the models
of Menou et al) and even then only fits for disks seen almost edge-on\footnote{Furthermore, even
the edge-on case may be ruled out in the case of moderate flaring 
as the disk would be expected to eclipse the star and
lower the optical luminosity -- Livio, Pringle \& Wood 2005.}.
Thus, passive black body models are ruled out.
Table~\ref{BigTab3} shows the
constraint on $\cos i$ we infer by requiring that the model flux be
less than the detected $8 \mu m$ level (assuming $a_i =10^{11} cm$).

\subsection{Dust only disks}

Chiang \& Goldreich (1997) introduced an extension to the standard disk
model incorporating a superheated dust layer in addition to the gaseous
disk. The temperature balance for the dust is different because the
emissivities in the UV/optical (where flux is absorbed) and in the
infrared (where flux is emitted) are markedly different for dust.
Thus, the disk can behave as if it is optically thick in absorption
but optically thin in emission. 

\begin{figure}
\begin{center}
\leavevmode
\includegraphics[width=14cm,angle=0]{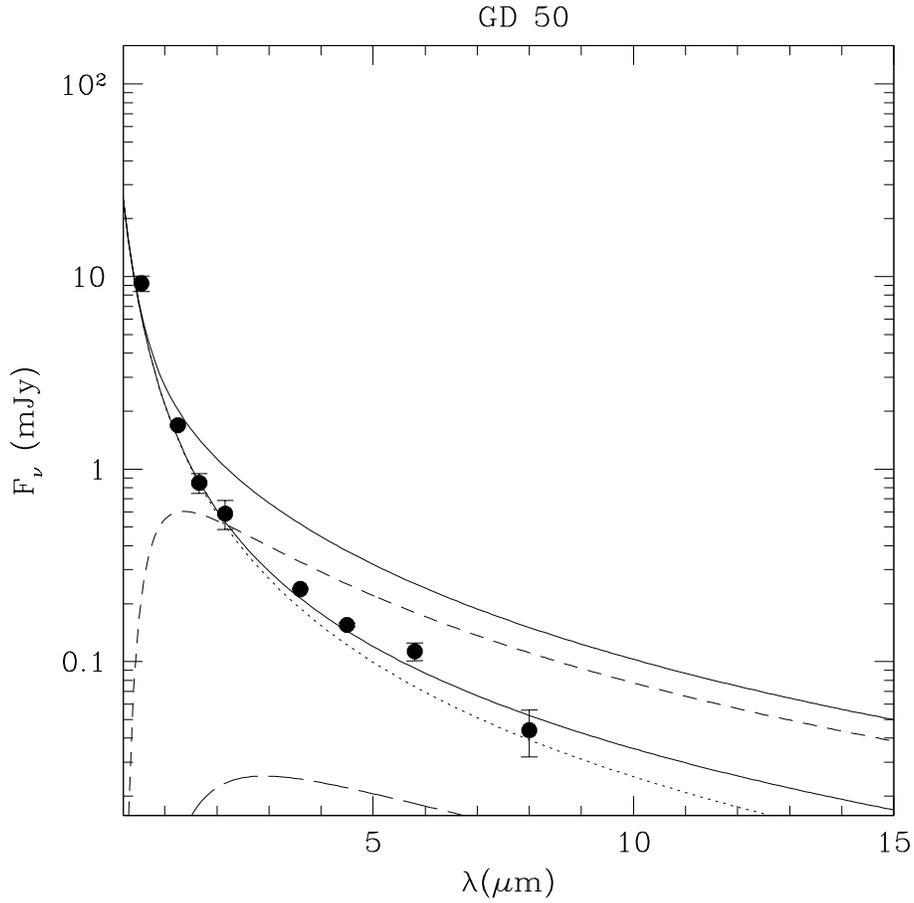}
\end{center}
\caption{ The solid points show the flux measurements for GD~50. The dotted
curve is a black body of the appropriate temperature. The short dashed curve
shows the emission from a flat disk composed of superheated dust, as described
in the text. Once again, the inclination is assumed to be 85$^{\circ}$. The
inner and outer radii are $10^{10}$cm and $10^{11}$cm respectively. The long
dashed curve shows the level of contribution allowed by the observations if
the dust disk is optically thin to incident radiation. The inclination in this
case is 45$^{\circ}$ and $\tau_{UV}=0.05$.
}
\label{Comp2}
\end{figure}

Given the fact that the disk in the situation under consideration can be almost entirely
metallic, it is of interest to consider the effects of a flat, dust-only, reprocessing
disk. Modelling this disk in the manner of Chiang \& Goldreich, we infer
a different temperature distribution from the previous section
$$
T = \left( \frac{R_*}{2 a} \right)^{\frac{2}{\beta + 4}} T_*
$$
where we have modelled the ratio of infrared to optical emissivities as
$\propto (T/T_*)^{\beta}$. We will assume $\beta \sim 1$ as in Chiang \& Goldreich.
 The emergent emission must also be diluted
by a factor $ 1 - exp(-\tau_{IR})$ where $\tau_{IR}$ is the optical
depth of the disk in the infrared. The resulting temperatures are somewhat
higher at a given radius than in the blackbody disk and this model is shown 
in Figure~\ref{Comp2}, as the short dashed line. In this instance,
varying the inner disk edge does not change the flux profile as
much as before, so we show only the case of $a_i = 10^{10}$cm.
This model is also ruled out as it requires all the observed systems
to be essentially edge on in order to meet the constraints.

If one reduces the mass of dust in the disk still further, the disk
becomes optically thin even in the ultra-violet. In this case, the absorbed
flux is reduced by a factor $1 - exp(-\tau_{UV})$, where $\tau_{UV}$ is
the average optical depth of the disk to incident stellar photons. This
additional parameter allows us to fix the disk inclination to more reasonable
values and determine the level of $\tau_{UV}$ needed to fit the data.
The long dashed curve in Figure~\ref{Comp2} shows this optically thin disk
with the same inner and outer edges as before, but now with an inclination
of $45^{\circ}$ and $\tau_{UV}=0.05$.

We can infer an approximate surface density $\Sigma \sim \tau_{UV}/\kappa_{UV}$
where $\kappa_{UV}$ is the appropriate opacity. For $\kappa_{UV} \sim 10^3 cm^2 g^{-1}$,
this means $\Sigma \sim 5 \times 10^{-5} g \, cm^{-2}$, which yields an approximate
disk mass
$$
M_{disk} \sim \pi R^2 \Sigma \sim 1.6 \times 10^{18} g < 10^{-15} M_{\odot}.
$$
This is the mass of an ordinary asteroid in the solar system -- a very stringent
limit on the amount of dust remaining in the system.

Table~\ref{BigTab3} also shows some quantitative limits for each system in the optically
thin limit. 
 We place constraints on $\tau_{UV}$ assuming an optically
thin, superheated dust disk inclined at $60^{\circ}$. One can also translate that into an estimate of the
dust mass.

\subsection{Planets}

We have thus far placed constraints on passively reprocessing dust disks.
If the material forms rapidly into planets, then the appropriate signature
is not that of a dust disk, but of a young, hot planet. However, such planets
will be almost impossible to detect. Since such planets would be made of
the same material as a white dwarf they obey roughly the same cooling
physics. Indeed, the young ages of the systems under discussion mean that
the physics is well approximated by the simple Mestel model (Mestel 1952;
see D'Antona \& Mazzitelli 1990 or Hansen 2004 for more extensive reviews).
In that model, luminosity at fixed age is proportional to mass, so that
the low mass planets will fade rapidly (much more rapidly than
a gaseous planet of similar mass) compared to the white dwarf and
will never become visible as an excess.

However, in the case of our two low-mass, Helium-core white dwarfs, there
is a possibility of finding brown dwarfs or normal planets as companions.

\subsubsection{Low Mass White Dwarfs}
\label{PlanetComps}

\begin{figure}
\begin{center}
\leavevmode
\includegraphics[width=14cm,angle=0]{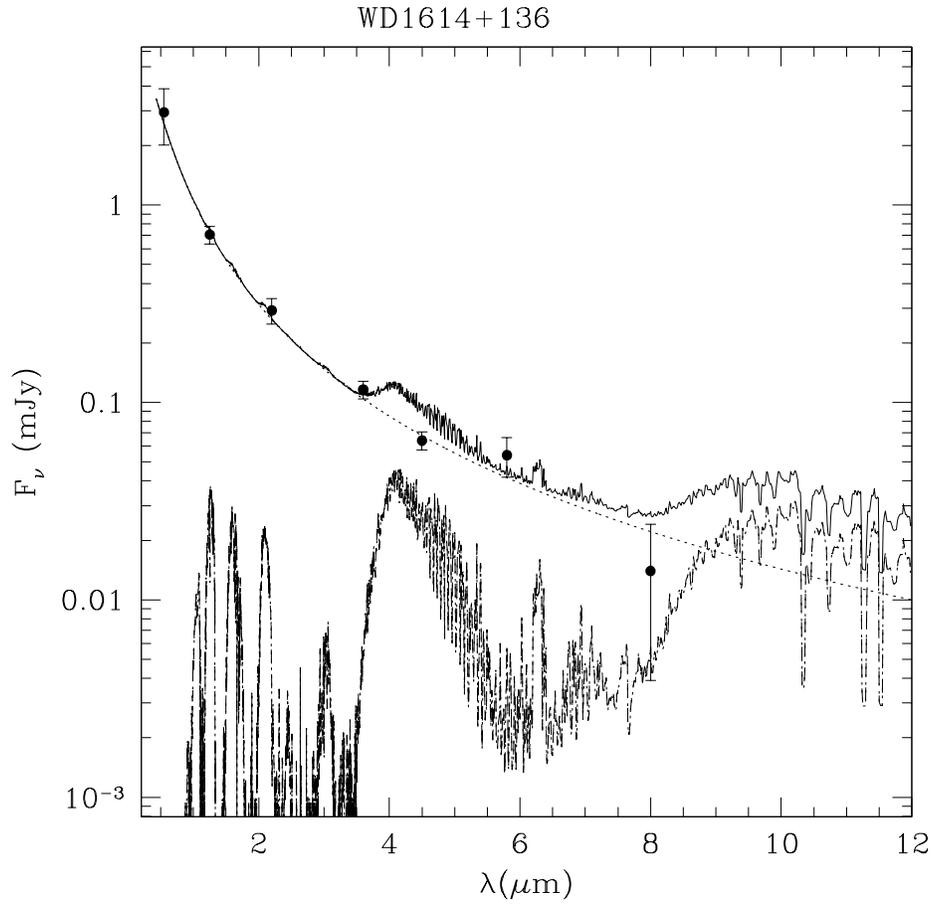}
\end{center}
\caption{ The dotted line is the white dwarf photosphere emission.
The dashed line shows the spectrum of a $10^8$~years old, $10 M_J$ 
brown dwarf and the solid line is the combined emission. The
model is ruled out (at the 2$\sigma$ level) by the photometry at
4.5$\mu m$.
}
\label{BD}
\end{figure}

 Two of our objects, WD1614+136 and WD1353+409, are low mass white dwarfs
without any obvious companion. It has been suggested that they are still
the product of truncated stellar evolution, but that the Roche Lobe overflow
was induced by a planet or a brown dwarf (Nelemans \& Tauris 1998). In this
case, we might expect the signature of a more traditional planetary mass
object. Figure~\ref{BD} shows the effect on the photometry of a 10$M_J$ brown dwarf
of age $10^8$~years (model taken from Burrows, Sudarsky \& Lunine 2003 -- BSL). 
This model is ruled out by the measurement in the IRAC2 band, which lies several
$\sigma$ below the predicted level. Lower mass or older objects are consistent
with this measurement.

\begin{figure}
\begin{center}
\leavevmode
\includegraphics[width=14cm,angle=0]{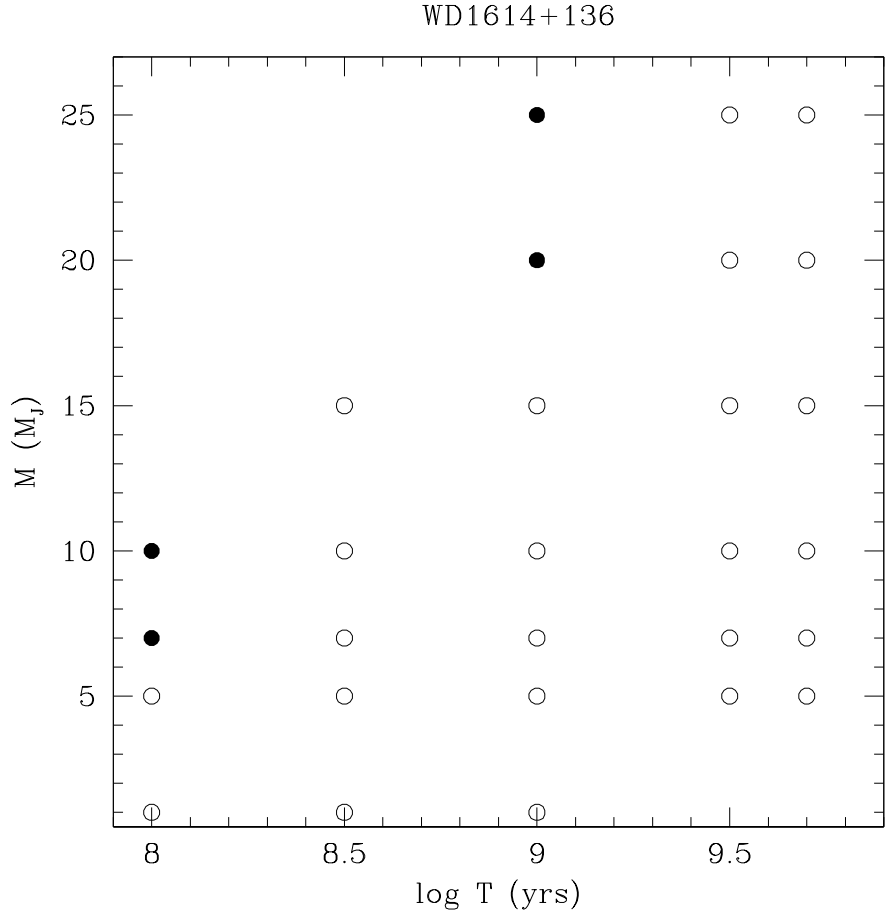}
\end{center}
\caption{ Solid points indicate models of planets from BSL that
could be detected by our observations if in orbit about WD1614+136.
In the case of open circles, these models would be undetectable
because they lie well below the photospheric emission. The cooling age
of the white dwarf is only $\sim 2 \times 10^7$~years, but if the planet
was formed at the same time as the original main sequence star it could
be considerably older.
}
\label{params}
\end{figure}

Figure~\ref{params} shows the various BSL models in terms of age
and mass. Solid points indicate models ruled out as companions to
WD1614+136 (in all cases it is because of an excess of 4.5$\mu m$ flux).
Open circles are models which are consistent with the observations.
We can rule out models down to $5 M_J$ for very young ages ($10^8$ years)
but cannot rule out models as massive as $25 M_J$ for ages $>10^9$ years.

Similar constraints are found for unresolved companions to WD1353+409.
If we consider the nearby ($\sim 5''$) object to be a physical companion,
then the flux ratio in the Rayleigh-Jeans tail is $\sim 4$ at short IRAC
wavelengths. This is consistent with the neighbour being a white dwarf
of similar radius but lower temperature ($\sim 5000$--6000~K), in a
 $\sim 630$~Au orbit. Increasing the radius to that of a brown dwarf
would imply temperatures $\sim 100$~K, which is not consistent with the
observed photometry (for either a black body or one of the BSL models). 
Of course, it could always be a background G or K dwarf of appropriate
temperature. However, for this line of sight, it would have to lie
$\sim 7$~kpc above the Galactic plane.

\section{Discussion}
\label{Discuss}

We have searched twelve hot, massive white dwarfs and two hot, undermassive ones, for signs of infrared
excesses indicative of a circumstellar dust disk reprocessing stellar radiation.
We have found no convincing evidence of any such excess. In all cases, the
objects are detected to wavelengths as long as $\lambda = 4.5 \mu m$ and
are found to be consistent with photospheric emission as predicted on the
basis of their optically determined temperatures and luminosities. In
several cases this conclusion can be extended out to $8 \mu m$. 

Our negative result can be used to place quantitative constraints on disk
models. For an optically thick, passively reprocessing blackbody disk, the
observations place stringent limits -- requiring either than the inclinations
be nearly edge on in all cases or that the inner radius of the disk be
located at distance $>10^{11}$cm (which is unlikely within the angular momentum 
constraints of the formation scenario). If we postulate a gas-free dust disk which is
optically thin, we can place constraints on the amount of dust mass present.
In all cases the mass cannot be larger than that of a 
solar system asteroid -- and most require $M < 10^{19}$g.

Our motivation for this search was that the mass distribution of white
dwarfs shows an excess of systems with masses $>1 M_{\odot}$. It has been
suggested that such an excess could result from the mergers of two
lower-mass white dwarfs driven by gravitational wave emission. If such
a merger leaves behind a circumstellar disk of tidally disrupted material,
one potential signature would be an infra-red excess resulting from
reprocessed stellar radiation.

Our stringent limits suggest that, if such disks do indeed form, the dust
must be removed or evolve on timescales shorter than the cooling times of
our sample white dwarfs ($\sim 10^7$~years). There are several possible
mechanisms. 

A dust disk does not remain in orbit indefinitely. Small grains are blown
out of the system by radiation pressure and larger grains spiral into the
star under the action of Poynting-Robertson drag. For our white dwarfs,
radiation pressure removes only the smallest grains, because even for
our hottest objects, $L \sim 0.1 L_{\odot}$. Poynting-Robertson drag is
significant, especially because our disks are somewhat more compact than
traditional protoplanetary disks. The inspiral time for dust under the action
of this drag is
$$
T_{PR} = \frac{4 \pi <s> \rho_s}{3} \frac{c^2 a^2}{L_*}
$$
where $<s>$ and $\rho_s$ are the average grain size and density respectively,
and $a$ is the distance from the star, of luminosity $L_*$. Numerically, this
is
$$
T_{PR} = 100 \, {\rm years} \left( \frac{<s>}{10 \mu m} \right) \left( \frac{\rho_s}{3 g cm^{-3}}
\right) \left( \frac{a}{10^{11} cm} \right)^2 \left( \frac{L_*}{10^{-2} L_{\odot}} \right)^{-1}.
$$
So, the removal time of dust in this system is very short.  Nevertheless, a
detectable signal is still expected if the dust is continuously replenished
by asteroid or comet collisions in the debris disk. Simple estimates of collisional
evolution in disks (e.g. Dominik \& Decin 2003) suggest that collisions between
asteroids occur on a characteristic timescale
$$
T_{coll} = 40 \, {\rm years} \left( \frac{a}{10^{11} cm} \right)^{3.5} 
\left( \frac{R}{1 km} \right) \left( \frac{M_{disk}}{10^{22} g} \right)^{-1}
$$
for a disk of total mass $M_{disk}$ composed of asteroids of size $R$. Thus, our
expectations correspond to a scaled-down version of a more traditional protoplanetary
debris system surrounding a main sequence star.

Nevertheless, we see no evidence of a dust signature -- if there was a merger
the dust was either removed very quickly before it could form larger bodies,
or else it formed sufficiently few large bodies that the collision time is 
long and the dust replenishment is low. Requiring therefore that $T_{coll}>T_{PR}$
and setting $R=4 km$ (so that the limits on the observable excess correspond to
the mass of a single disrupted asteroid), places a limit on the characteristic
total disk mass in this simple model $M_{disk} < 10^{21} g$. One can potentially
raise this limit by a factor $\sim 1/Z_{\odot} \sim 50$ if one postulates that
only refractory dust can survive at the temperatures inferred for these compact
disks. If most of the Carbon and Oxygen that emerges from the disrupted white
dwarf forms ices which are rapidly sublimated, then the amount of dust is limited
by the available refractory elements. However, the resulting limit is still no larger
than an asteroid.

The fact that disks can indeed occur around white dwarfs and that they can be
detected is shown by the infra-red excesses of the stars G~29-38 (Zuckerman \& Becklin 1987) and 
GD~362 (Becklin et al 2005; Kilic et al 2005). These white dwarfs are also DAZ
stars, which show evidence for trace amounts of Calcium and other refractory
elements in their atmospheres (Zuckerman \& Becklin 1987; Gianninis et al 2004). 
The disks in these cases are thought to be the products of the disruption of an 
asteroid (e.g. Jura 2003)
rather than debris from a merger (while GD~362 is indeed quite massive, G29-38 has
a very ordinary mass). Nevertheless, this demonstrates that the basic model we
set out to test is viable and that our null result is significant.

It is, of course, also quite possible that there are one or more flaws in the
merger model for the origin of massive white dwarfs. One possibility is if
 one of the two merging stars has a helium core. If this were the case,
then the accretion of the helium onto the surface of the merger product produces
a 
helium burning star rather than a star+disk system (Iben 1990), However,
this eventuality seems unlikely for the massive systems considered here, because that requires
that one of the components be $<0.4 M_{\odot}$ and therefore the primary
would have to be massive to begin with.

Finally, the massive white dwarfs may simply be the result of single star
evolution after all.
LBH confirm a significant excess of massive white dwarfs in the sample
of older, cooler white dwarfs drawn from the Palomar Green (PG) sample (Green, Schmidt \& Liebert 1986). Indeed,
they assert that 80\% of the white dwarfs in the PG sample with mass $> 0.8 M_{\odot}$ 
are formed by some other process other than single star evolution. 
Nevertheless, they also note that the fraction of massive ($>0.8 M_{\odot}$) white dwarfs
with $M > 1 M_{\odot}$ is higher in the UV/X-ray selected surveys than in the PG
survey. This might be due to the fact that the former are selected from an
all-sky survey while the latter is restricted to high Galactic latitudes. 
LBH suggest that the EUVE samples may have a higher fraction of massive
white dwarfs born from single star evolution than the PG sample because the
O/B stars in Gould's belt may provide a significant enhancement in the
current local birthrate of such objects. Thus, it is still not entirely clear
how much of a binary contribution is expected to contribute to our sample (drawn
from the EUV/X-Ray catalogues to get the hottest and so youngest systems).
Finally, the constraints on the true shape of the initial--final mass relation
for stars are still rather weak and it is possible (although far from proven)
that the features in the mass distribution are simply the result of a non-linear
relation (e.g. Ferrario et al 2005) between the main sequence and white dwarf masses of stars.

In conclusion, we have examined 14  hot white dwarfs for signs of
near-infrared excess. These have all been suggested as
potential merger remnants (12 supermassive white dwarfs and 2 undermassive white dwarfs).
In no case do we find any evidence of infrared excess.

\acknowledgements
This work is based on observations made with the {\em Spitzer Space Telescope},
which is operated by the Jet Propulsion Laboratory (JPL), California Institute
of Technology (CIT) under National Aeronautics and Space Administration (NASA)
contract 1407. We thank NASA, JPL and the {\em Spitzer} Science Center for
support through Spitzer contracts 1264152 to UCLA.
BH acknowledges the support of an Alfred P. Sloan Foundation Fellowship
and thanks Mike Jura, Jim Liebert, Ben Zuckerman and Jay Farihi for comments.
This publication makes use of data products from the Two-Micron All Sky
Survey, which is a joint project of the University of Massachusetts and the
Infrared Processing and Analysis Center/CIT, funded by NASA and the
National Science Foundation. This publication has also made use of the
online McCook \& Sion White Dwarf Catalog and NASA's Astrophysics Data System.

\clearpage

\clearpage

\begin{deluxetable}{lllll}
\tablecolumns{7}
\tablewidth{0pc}
\tablecaption{Sample Characteristics \label{BigTab1}}
\tablehead{\colhead{Name}  & \colhead{$T_{\rm eff}$} & \colhead{$\log g$} & \colhead{M}
& \colhead{d}  \\
 & (K) &  & ($\rm M_{\odot}$) & (pc)   }
\startdata
EUVE0138+25.3/WD0136+251 & 39400(200)\tablenotemark{a}  & 9.12(13)\tablenotemark{a}  & 1.32(8) & 76  \\
EUVE0237-12.3/WD0235-12.5 & 32400(200)\tablenotemark{a} & 8.64(4)\tablenotemark{a} & 1.03(2) & 66  \\
EUVE0443-03.7/WD0440-038 & 65140(200)\tablenotemark{a} & 9.12(12)\tablenotemark{a} & 1.33(7)  & 144   \\
EUVE0521-10.4/WD0518-105 & 33000(300)\tablenotemark{a} & 8.70(8)\tablenotemark{a} & 1.07(5) & 99  \\
EUVE0916-19.7 & 56400(2600)\tablenotemark{a} & 9.12(20)\tablenotemark{a} & 1.33(11) & 164 \\
EUVE1659+44.0/WD1658+441 & 30510(200)\tablenotemark{b} & 9.36(7)\tablenotemark{b} & 1.41(4)  & 27 \\
EUVE1024-30.3/WD1022-301 & 34800(500)\tablenotemark{c} & 9.20(12)\tablenotemark{c} & 1.27(3) & 61 \\
REJ0003+43/WD0001+433 & 42400(1200)\tablenotemark{a} & 9.30(12)\tablenotemark{a} & 1.37(6) & 101 \\
REJ0348-005/WD0346-011 & 43200(500)\tablenotemark{a} & 9.21(5)\tablenotemark{a} & 1.37(3)  & 29 \\
REJ1746-706/WD1740-706 & 46800(800)\tablenotemark{c}  & 8.94(8)\tablenotemark{c} & 1.18(3) & 79 \\
EUVE0317-85.5/WD0325-857 & 40000(10000) & 9.45(15) & 1.33(3) & 36 \\
WD1614+136 & 22400\tablenotemark{d} & 7.34\tablenotemark{d} & 0.33 & 110 \\
WD1353+409 & 23600\tablenotemark{d} & 7.54\tablenotemark{d} & 0.40 & 126\\
EUVE0653-164/WD0652-563 & 35200(600)\tablenotemark{a} & 8.88(10)\tablenotemark{a} &  1.18(6) & 107  \\
\enddata
\tablenotetext{a}{from Vennes et al 1997}
\tablenotetext{b}{from Schmidt et al 1992}
\tablenotetext{c}{from Marsh et al 1997}
\tablenotetext{d}{from Bergeron et al 1992}
\tablecomments{The masses are derived from models with hydrogen atmospheres.
Approximate distances are derived by matching the absolute magnitude derived from
the spectral fits to the apparent magnitude.}
\end{deluxetable}

\begin{deluxetable}{lllcccc} 
\tablecolumns{7} 
\tablewidth{0pc} 
\tablecaption{IR Fluxes of Hot White Dwarfs \label{BigTab2}}
\tablehead{\colhead{Name}  & \colhead{m$_V$} & \colhead{m$_J$} & \colhead{F$_{3.6}$}
& \colhead{F$_{4.5}$} & \colhead{F$_{5.8}$} & \colhead{F$_{8}$} \\
 & & & (mJy) & (mJy) & (mJy) & (mJy)  }
\startdata 
EUVE0138+25.3/WD0136+251 & 15.87(3)  & 16.48(6) & 0.047(2) & 0.027(3) & $<0.012$ & $<0.012$  \\
EUVE0237-12.3/WD0235-12.5 & 15.4(1)& 15.74(6)\tablenotemark{b} & 0.117(2) & 0.075(3) & 0.043(11) & 0.047(10) \\
  &  &  & 0.107(1) & 0.065(2) & 0.040(9) & 0.031(8) \\
EUVE0443-03.7/WD0440-038 & 16.7(2) & 17.38(5)\tablenotemark{b} & 0.019(2) & 0.022(2) & $<0.011$ & $<0.011$  \\
EUVE0521-10.4/WD0518-105 & 15.82(3) & 16.76(20)\tablenotemark{b}  & 0.052(2) & 0.037(3) & 0.011(10) & $<0.011$  \\
EUVE0916-19.7 & 17.3 & 13.94(2)\tablenotemark{b} & 0.020(2) & 0.011(3) & 0.018(11) & $<0.011$ \\
EUVE1659+44.0/WD1658+441 & 14.62 & 15.47(10)\tablenotemark{b} & 0.129(2) & 0.078(2) & 0.039(10) & $<0.009$ \\
EUVE1024-30.3/WD1022-301 & 16.7(3) & 16.97(8) & 0.037(2) & 0.024(3) & $<0.011$ & $<0.010$ \\
REJ0003+43/WD0001+433 & 16.8(3) & 17.0(1) & 0.020(2) & 0.014(3) & $<0.011$ & 0.020(10) \\
REJ0348-005/WD0346-011 & 14.04(2) & 14.97(1) & 0.238(2) & 0.155(3) & 0.113(12) & 0.044(12) \\
REJ1746-706/WD1740-706\tablenotemark{a} & 16.6(3) & 17.2(1) &0.031(1) & 0.015(1) & 0.009(6) & 0.010(6) \\
EUVE0317-85.5/WD0325-857 & 14.8 & 15.43(2) & 0.156(2) & 0.094(2) & 0.050(10) & 0.029(10) \\
WD1614+136 & 15.24 & 15.84(2) & 0.116(2) & 0.065(2) & 0.053(10) & 0.014(10) \\
WD1353+409 & 15.55 & 16.20(3) & 0.080(2) & 0.052(3) & 0.049(9) & 0.035(9) \\
EUVE0653-164/WD0652-563 & 16.4 & 17.4(1) & 0.019(2) & 0.018(2) & 0.023(11) & 0.022(10) \\
\enddata 
\tablenotetext{a}{The exposure length for this object was twice as long as for the others.}
\tablenotetext{b}{taken from Green, Ali \& Napiwotzki 2000.}
\tablecomments{ The quoted IRAC errors are statistical. An additional calibration uncertainty
of $\sim 2\%$ (Fazio et al 2004) is applicable.}
\end{deluxetable} 
 
\begin{deluxetable}{lllccc}
\tablecolumns{7}
\tablewidth{0pc}
\tablecaption{Constraints on Disks \label{BigTab3}}
\tablehead{\colhead{Name}  & \colhead{D} & \colhead{F$_8$} & \colhead{$\cos i$}
& \colhead{$\tau_{UV}$} & \colhead{M$_{20}$} \\
 & (pc) & (mJy) &  &  & (g)  }
\startdata
EUVE0138+25.3/WD0136+251 & 76 & $<0.012$ & $<0.024$ & $<0.10$ & $<0.03$ \\
EUVE0237-12.3/WD0235-12.5  & 66  & $<0.012$ & $<0.07$ & $<0.30$ & $<0.10$ \\
EUVE0443-03.7/WD0440-038 & 144 & 0.047 & $<0.03$ & $<0.34$ & $<0.11$  \\
EUVE0521-10.4/WD0518-105 & 99 & $<0.011$ & $<0.04$ & $<0.16$ & $<0.05$ \\
EUVE0916-19.7 & 164 & $<0.011$ & $<0.10$ & $<0.44$ & $<0.14$ \\
EUVE1659+44.0/WD1658+441
 & 27 & $<0.011$ & $<0.002$ & $<0.010$ & $<0.003$ \\
EUVE1024-30.3/WD1022-301
 & 61
& $<0.009$ & $<0.017$ & $<0.06$ & $<0.02$ \\
REJ0003+43/WD0001+433 & 101 & 0.020 & $<0.07$ & $<0.30$ & $<0.09$ \\
REJ0348-005/WD0346-011 & 29 & 0.044 & $<0.013$ & $<0.05$ & $<0.02$ \\
REJ1746-706/WD1740-706 & 79 & 0.010 & $<0.022$ & $<0.09$ & $<0.03$ \\
EUVE0317-85.5/WD0325-857 & 36 & 0.029 & $<0.013$ & $<0.06$ & $<0.02$ \\
WD1614+136
 & 110 & 0.014 & $<0.060$ & $<0.25$ & $<0.08$ \\
WD1353+409 & 126 & 0.035 & $<0.200$ & $<0.843$ & $<0.265$ \\
EUVE0653-164/WD0652-563 & 107 & 0.022 & $<0.09$ & $<0.37$ & $<0.12$ \\
\enddata
\tablecomments{Constraints on inclination for a flat, blackbody disk
and on mass for an optically thin dust disk.
}
\end{deluxetable}

\end{document}